\documentclass[twoside,fleqn]{article}
\usepackage{amssymb}
\usepackage{amsbsy}
\usepackage{amsfonts}
\usepackage{epsfig}
\usepackage{espcrc2}

\def\thebibliography#1{\section*{References}\list{\arabic{enumi}.}
  {\settowidth\labelwidth{#1.}\leftmargin=1.67em
   \labelsep\leftmargin \advance\labelsep-\labelwidth
   \itemsep0pt \parsep0pt
   \usecounter{enumi}}\def\makelabel##1{\rlap{##1}\hss}%
   \def\newblock{\hskip 0.11em plus 0.33em minus -0.07em}
   \sloppy \clubpenalty=4000 \widowpenalty=4000 \sfcode`\.=1000\relax}

\title{\vspace{-2.0cm}
       {\normalsize DESY 00--161}    \\[-0.2cm]
       {\normalsize November 2000}   \\[1.25cm]
        Is the chiral U(1) theory trivial?$\,$\thanks{Talk given by 
G. Schierholz at {\it Lattice 2000}, Bangalore, India}}

\author{V. Bornyakov\address{Institute for High Energy Physics IHEP, 
        RU-142284 Protvino, Russia}, 
        A. Hoferichter\address{Deutsches Elektronen-Synchrotron DESY, John von
        Neumann-Institut f\"ur Computing NIC,\\ 
        \hspace{0.165cm}D-15735 Zeuthen, Germany}, and
        G. Schierholz$^{\rm b,}$\address{Deutsches Elektronen-Synchrotron DESY,
        D-22603 Hamburg, Germany}}

\begin{document}

\begin{abstract}
\vspace*{0.25cm}
The chiral U(1) theory differs from the corresponding vector theory by an
imaginary contribution to the effective action which amounts to a phase factor
in the partition function. The vector theory, {\it i.e.} QED, is known to be 
trivial in the continuum limit. It is argued that the presence of the phase 
factor will not alter this result
and the chiral theory is non-interacting as well. 
\vspace*{0.25cm}
\end{abstract}

\maketitle

\renewcommand{\thefootnote}{\fnsymbol{footnote}}

\section{Introduction}

In the continuum limit QED appears to be a trivial theory of massless 
fermions. Vice versa, massless fermions turn out to be
non-interacting irrespective of the value of the cut-off~\cite{QED}. The 
chiral theory, {\it i.e.} the theory of charged massless left- and 
right-handed fermions interacting via photon exchange, differs from the 
vector theory by an imaginary contribution to 
the effective action, while the real part of the effective action is 
vector-like~\cite{AG}. This raises the
question as to whether the imaginary part can turn a non-interacting theory 
into an interacting one. If not, that has interesting consequences. 

The presence of the imaginary part is the main obstacle in simulating
chiral fermions on the lattice. In this talk we shall follow the `continuum
fermion approach' (CFA)~\cite{CFA} to the problem. The idea here is to compute
the fermion action in the continuum. Starting from a lattice of extent $L$
with spacing $a$, the original lattice on which one does the 
simulations, one 
constructs a finer lattice with spacing
$a_f$ on which one puts the fermions. The action for a 
single fermion of charge $e_\alpha$ (in units of $e=1/\sqrt{\beta}$) and
chirality $\epsilon_\alpha = \pm 1$ is taken to be  
\newpage
\vspace*{-1.0cm}
\begin{eqnarray}
\label{action}
S_{e_\alpha,\epsilon_\alpha} \!\!\!\!\!\!&=&\!\!\!\! 
\frac{1}{2a_f}\sum_{n,\mu} 
\bar{\psi}(n) \gamma_{\mu} \Big\{\big[P_{-\epsilon_\alpha}
+P_{\epsilon_\alpha}(U_{\mu}(n))^{e_\alpha}\big] \nonumber \\
\!\!\!\!\!\!&\times&\!\!\!\! \psi(n+\hat{\mu}) - \big[P_{-\epsilon_\alpha} 
+ P_{\epsilon_\alpha}(U^{\dagger}_{\mu}(n-\hat{\mu}))^{e_\alpha}\big] 
\nonumber \\
\!\!\!\!\!\!&\times&\!\!\!\! \psi(n-\hat{\mu})\Big\} + S_W, 
\end{eqnarray}
where $U_{\mu}=\exp({\rm i}A_\mu) \in {\rm U(1)}$ and 
$P_{\epsilon_\alpha} = (1 +\epsilon_\alpha \gamma_5)/2$, and $S_W$ is the
ungauged Wilson term:
$$
S_W = \frac{1}{2a_f}\sum_{n,\mu} \bar{\psi}(n) 
\big\{2\psi(n)-\psi(n+\hat{\mu})-\psi(n-\hat{\mu})\big\}.
$$
The action (\ref{action}) obeys the Golterman-Petcher shift symmetry so 
that in the 
limit $a_f \rightarrow 0$ it describes one interacting massless fermion of 
chirality $\epsilon_\alpha$, and one free fermion of chirality 
$-\epsilon_\alpha$ which decouples from the system. 
The lattice effective action $W_{e_\alpha,\epsilon_\alpha}$ (for finite $a_f$)
follows from 
\begin{equation}
\exp(-W_{e_\alpha,\epsilon_\alpha}) =\int
{\mathcal{D}}\bar{\psi}{\mathcal{D}}\psi  
\exp(-S_{e_\alpha,\epsilon_\alpha}),
\end{equation}
and the continuum action is given by  
\begin{equation}
\widehat{W}_{e_\alpha,\epsilon_\alpha} = \lim_{a_f \rightarrow 0}
(W_{e_\alpha,\epsilon_\alpha} + C),
\end{equation}
where $C$ is a local bosonic counterterm whose purpose is to render 
${\rm Re}\,\widehat{W}_{e_\alpha,\epsilon_\alpha}$ invariant under chiral 
gauge transformations. The counterterm is known analytically.
For the real part of the effective action we obtain, in agreement 
with the continuum result~\cite{AG}, 
\begin{equation} \label{re}
{\rm Re}\: \widehat{W}_{e_\alpha,\epsilon_\alpha} = 
\frac{1}{2} \big(\widehat{W}_{e_\alpha,\, V} + \widehat{W}_{0} \big),
\end{equation}
where $\widehat{W}_{e_\alpha, \, V}$ and $\widehat{W}_{0}$ are the actions of
the corresponding vector theory and the free theory,
respectively, in the $a_f \rightarrow 0$ limit.
For the imaginary part we obtain
\begin{equation}
{\rm Im}\: \widehat{W}_{e_\alpha,\epsilon_\alpha} = 
{\mathcal A} + \pi \epsilon_\alpha \eta_{e_\alpha}\:[{\rm mod}\, 2\pi] ,
\end{equation}
where ${\mathcal A}$ is the anomalous part of the effective action, and 
$\eta_{e_\alpha}$, the so-called $\eta$ invariant, is a gauge invariant 
quantity. The anomaly cancelling condition is
\begin{equation}
\sum_\alpha \epsilon_\alpha e_{\alpha}^3 = 0.
\end{equation}

\begin{figure}[tbp]
\vspace*{-0.25cm}
  \begin{center}
    \epsfig{file=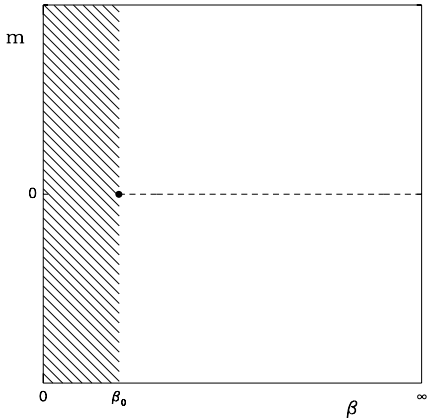,height=6.0cm,width=6.0cm}
\vspace*{-0.75cm}
\caption{The phase diagram of the vector theory in terms of bare 
coupling and bare mass. The horizontal line is a line of second
order phase transitions, which ends in a (tri-)critical point. In the
shaded area the compact and the non-compact theory differ. In the compact 
theory $\beta_0 \approx 1.$}
    \label{fig1}
  \end{center}
\vspace*{-0.85cm}
\end{figure}

The vector theory has been extensively studied
for compact~\cite{Andreas} and non-compact~\cite{QED} U(1) fields. The phase 
diagram is shown in Fig.~1. 
Common to both formulations is that the theory has a line of second order
phase transitions at zero bare mass, extending from (some) $\beta_0 > 0$ to 
$\beta = \infty$, on which chiral symmetry is restored and the theory is 
massless. It is this second order line on which one can take the (quantum)
continuum limit. In the non-compact case this has been shown~\cite{QED}
to correspond to a non-interacting theory, and it is believed that the same 
is true for the compact theory. The chiral theory would naturally live on the  
second order line. But we know of cases where a phase factor changes the 
properties of the theory.

We consider compact gauge fields. In the absence of monopoles
the gauge fields decompose into sectors of integer magnetic flux
$m_{\mu\nu}$ and topological 
charge~\cite{Phillips} $Q=\frac{1}{8}\epsilon_{\mu\nu\rho\sigma}
m_{\mu\nu}m_{\rho\sigma}$. The configurations with non-vanishing flux are 
strongly suppressed, however, so that we do not expect to see any differences 
between the compact and the non-compact formulation of the theory in the 
chirally symmetric phase. 

\section{Im $\boldsymbol{W}$: does it make a difference?}

As a first test of the method we have computed the anomaly ${\mathcal A}$ for 
a set of plane wave potentials. This allowed us to do calculations on (fine) 
lattices as large as $L_f = 16$ and to compare our results with the 
continuum expression given by the triangle diagram in Fig.~2. We found 
excellent agreement between the numerical 
results extrapolated to $L_f \rightarrow \infty$ and the analytical values.  

\begin{figure}[tbp]
  \begin{center}
    \epsfig{file=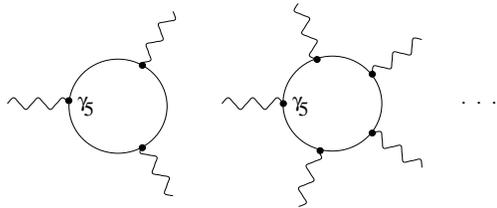,height=2.75cm,width=6.5cm}
\vspace*{-0.5cm}
\caption{Perturbative expansion of the imaginary part of the effective action:
  the triangle (three-leg) diagram and higher order contributions.}
\vspace*{-1.0cm}
  \label{fig2}
  \end{center}
\end{figure}

We now turn to the $\eta$ invariant.~\footnote{In the chiral 
Schwinger model~\cite{BB} $\eta$ receives contributions from toron fields 
$t_\mu$ only and assumes values between $- 1$ and $+ 1$, no matter how
small $L$ is. Furthermore, it is 
non-analytic at $t = 0$ and has no perturbative expansion.} 
We consider three different charges, $e_\alpha=1/2$, 1 and 2. The calculation
proceeds from $L=4$ (original) lattices. The largest fine lattice we were able
to explore was $L_f \equiv (a/a_f)L=8$. We use the Lanczos algorithm to 
compute the fermion
determinant. Because the fermion matrix is non-hermitean, complete 
re-orthogonalization of the Lanczos vectors is necessary. This makes the
memory demand grow like $L_f^8$, which limited our calculations so far. 
The fine lattice is
constructed as follows. First we interpolate the original lattice gauge field 
to the continuum~\cite{GKSW}. Out of the continuum gauge field we then 
construct a lattice gauge field on the fine lattice. The origin of the fine 
lattice is chosen at random, each choice corresponding to a different
interpolation. This introduces an error. The error can be computed from 
sampling the action over a set of interpolations. This error vanishes as 
$a_f \rightarrow 0$. 

\begin{figure}[tbp]
  \begin{center}
    \epsfig{file=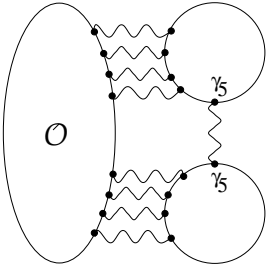,height=3.5cm,width=3.5cm}
\vspace*{-0.5cm}
\caption{A perturbative contribution to the connected correlation function
$\langle {\mathcal O} \cos\pi\eta_a \rangle_{\rm Re}^{\rm con}$ in
(\ref{corr}), where ${\mathcal O}$ is pictured as a fermion loop.} 
\vspace*{-1.1cm}
  \label{fig3}
  \end{center}
\end{figure}

We have chosen to work at $\beta = 2$. This coupling lies well 
in the symmetric phase of the vector model where we may hope to find massless 
fermions. The Monte Carlo sampling proceeds in two steps. In the first step 
we generate quenched gauge field configurations. On these configurations we 
then compute the fermion determinant which we include in the observable. 
This is justified if the action does not fluctuate too much. We
have checked that (\ref{re}) is 
fulfilled. In Table 1 we present the real and imaginary part of the effective
action of the anomaly-free model $\epsilon_\alpha e_\alpha = +1, \alpha = 1,
\cdots, 8$, $\epsilon_\alpha e_\alpha = -2, \alpha = 9$ for 6 consecutive 
gauge field configurations and for various levels of interpolation. We also 
show the extrapolation to $L_f = \infty$. The imaginary part of the effective
action turns out to be surprisingly small, and it appears that the larger the
value is the larger is the corresponding real part which suppresses its effect
even further.

\begin{table*}[tbh]
\begin{center}
\begin{tabular}{|c|c|l|l|l|l|l|l|} \hline
& & \multicolumn{6}{|c|}{$L_f$} \\ \cline{3-8}
& $\#$ & \multicolumn{1}{|c|}{4}
& \multicolumn{1}{|c|}{5} & \multicolumn{1}{|c|}{6} & \multicolumn{1}{|c|}{7}
& \multicolumn{1}{|c|}{8} & \multicolumn{1}{|c|}{$\infty$} \\  
\hline 
&1 &17.3(25)
&17.7(02)
&17.7(18)
&17.9(01)
&19.8(36)
&18.1(03)
  \\ 
&2 &13.9(18)
&15.2(56)
&15.2(04)
&15.5(02)
&18.1(37)
&16.3(08)
  \\ 
${\rm Re}\,\widehat{W}_a$ &3 &17.1(17)
&17.2(02)
&17.1(03)
&17.2(01)
&19.1(29)
&17.2(03)
  \\ 
&4 &16.5(17)
&17.3(01)
&17.7(04)
&18.2(01)
&19.9(51)
&19.1(02)
 \\ 
&5 &23.3(19)
&24.1(02)
&24.9(04)
&25.3(02)
&27.2(34)
&26.5(04)
  \\  
&6 &25.2(18)
&27.1(158)
&31.8(363)
&76.0(1100)
&25.3(26)
&25.4(35)
  \\ \hline 
&1 &0.0001(6)
&0.0005(6)
&0.0010(7) &0.0013(5) 
&0.0014(2)
&0.0019(3)
  \\ 
&2 &\hspace*{-0.12cm}-0.0008(3)
&\hspace*{-0.12cm}-0.0015(5)&\hspace*{-0.12cm}-0.0020(4)
&\hspace*{-0.12cm}-0.0021(5)
&\hspace*{-0.12cm}-0.0019(1) & \hspace*{-0.12cm}-0.0023(1)  \\ 
$\pi \eta_a$ &3 &\hspace*{-0.12cm}-0.0029(8)
&\hspace*{-0.12cm}-0.0045(6)
&\hspace*{-0.12cm}-0.0039(3)
&\hspace*{-0.12cm}-0.0022(4)
&\hspace*{-0.12cm}-0.0002(3) &\hspace*{-0.12cm}-0.0001(4)  \\ 
&4 &\hspace*{-0.12cm}-0.0012(3)
&\hspace*{-0.12cm}-0.0035(4) 
&\hspace*{-0.12cm}-0.0060(7) 
&\hspace*{-0.12cm}-0.0088(3) &  
\hspace*{-0.12cm}-0.0104(8)  &\hspace*{-0.12cm}-0.0120(4) \\ 
&5 &\hspace*{-0.12cm}-0.0016(10)\hspace*{-0.12cm}
&\hspace*{-0.12cm}-0.0069(23)\hspace*{-0.12cm} 
&\hspace*{-0.12cm}-0.0295(34)\hspace*{-0.12cm} 
&\hspace*{-0.12cm}-0.0676(102)\hspace*{-0.12cm} 
&\hspace*{-0.12cm}-0.0805(51)\hspace*{-0.12cm}  
&\hspace*{-0.12cm}-0.0604(37)\hspace*{-0.12cm} \\  
&6 &0.0063(10)\hspace*{-0.12cm}
&0.0146(26)\hspace*{-0.12cm} 
&0.0207(23)\hspace*{-0.12cm} 
&0.0224(15) &  
0.0225(10)\hspace*{-0.12cm}  & 0.0287(12)\hspace*{-0.12cm}  \\ \hline 
\end{tabular}
\vspace*{0.4cm}
\caption{The real and imaginary part of the effective action of the 
anomaly-free model for 6 consecutive gauge field configurations. The errors 
represent the ambiguity in the interpolation.}    
\vspace*{-0.6cm}
\end{center}
\end{table*}

With the imaginary part being so small, the first 
question which comes to mind is whether ${\rm Im}\,\widehat{W}$ can be 
described by perturbation theory. The first few diagrams are given in Fig.~2.
If so, we would expect to find for a given $\epsilon_\alpha$
\begin{equation}
\begin{tabular}{llrcrcrc}
$\!\!\!e_\alpha=\frac{1}{2}: $&$\!\!\! {\rm Im}\,\widehat{W} =$&$\!\!\! 
\frac{1}{8} A \!\!\!$ &$\!\!\!+\!\!\!$& 
$\!\!\!\frac{1}{32} B\!\!\!$ &$\!\!\!+\!\!\!$& $\!\!\!\frac{1}{128} C \!\!\!$
&$\!\!\! + \cdots$, \\
$\!\!\!e_\alpha=1: $&$\!\!\! {\rm Im}\,\widehat{W} =$&$\!\!\! A \!\!\!$ &
$\!\!\!+\!\!\!$& 
$\!\!\! B\!\!\!$ &$\!\!\!+\!\!\!$& $\!\!\! C\!\!\!$&$\!\!\! + \cdots$, \\
$\!\!\!e_\alpha=2: $&$\!\!\! {\rm Im}\,\widehat{W} =$&$\!\!\! 8 A\!\!\!$ 
&$\!\!\!+\!\!\!$& 
$\!\!\! 32 B\!\!\!$ &$\!\!\!+\!\!\!$& $\!\!\! 128 C\!\!\!$&$\!\!\! + \cdots$, 
\end{tabular}
\end{equation}
where $A$, $B$ and $C$ are the contributions from the 3-, 5- and 7-leg 
diagrams, respectively. We find indeed good agreement with the perturbative 
behavior with $|A|$ : $|B|$ : $|C|$ $\sim$ 1 : 0.1 : 0.01.~\footnote{This
behavior is quite different from what we found in the chiral Schwinger model.} 
We expect to obtain similar ratios on larger lattices.

Let us consider an observable ${\mathcal O}$ of even parity now. Its
expectation value can be written  
\begin{eqnarray}
\langle {\mathcal O}\rangle \!\!\!\!\!&=&\!\!\!\! \frac{\langle 
{\mathcal O}\mbox{e}^{{\rm i}
\pi\eta_a} \rangle_{\rm Re}}{\langle \mbox{e}^{{\rm i} \pi\eta_a} 
\rangle_{\rm Re}} \nonumber \\
\!\!\!\!\!&=&\!\!\!\! \langle {\mathcal O}\rangle_{\rm Re} 
\!+\!\frac{\langle {\mathcal O} 
\cos\pi\eta_a \rangle_{\rm Re} \!- \langle {\mathcal O}\rangle_{\rm Re} 
\langle \cos\pi\eta_a \rangle_{\rm Re}}
{\langle \cos\pi\eta_a \rangle_{\rm Re}} \nonumber \\
\!\!\!\!\!&=&\!\!\!\! \langle {\mathcal O}\rangle_{\rm Re}
\!+\! \langle {\mathcal O} \cos\pi\eta_a \rangle_{\rm Re}^{\rm con},
\label{corr}
\end{eqnarray}
where `Re' means that the path integral is done over the real part of the 
effective action only, which corresponds to the vector theory given by
the action (\ref{re}) with 4 + $x$ active flavors, 
and `con' refers to the totally connected correlation function. 
In Fig.~3 we show a corresponding diagram. 
Using Schwarz's inequality and the results in Table 1 we estimate  
\begin{eqnarray}
|\langle {\mathcal O}\rangle - \langle {\mathcal O}\rangle_{\rm Re}| &\leq& 
\sigma_{\rm Re}({\mathcal O}) 
\frac{\sigma_{\rm Re}(\cos\pi\eta_a)}{\langle \cos\pi\eta_a \rangle_{\rm Re}}
\label{estimate1} \\[0.5em]
&\approx& \sigma_{\rm Re}({\mathcal O}) \times 3\:\cdot 10^{-5},
\label{estimate2}
\end{eqnarray}
where $(\sigma_{\rm Re}({\mathcal O}))^2$ is the variance of the operator 
${\mathcal O}$. A similar estimate is found for 
the parity-odd operator.
This indicates that the chiral U(1) theory is basically vector-like. We 
do not expect that this estimate will change significantly on larger volumes. 
The connected correlation function in (\ref{corr}) stays constant 
as $L \rightarrow \infty$ in any order of perturbation theory. And studies of
the vector theory~\cite{eos} have shown that finite size effects vanish
proportional to the fermion mass $m$ in the chiral limit.

At small $m$ the vector theory was well described by renormalized
perturbation theory~\cite{QED} with $e_R^2 \propto 1/\ln(m_R)$ as $m_R
\rightarrow 0$, $e_R$ ($m_R$) being the renormalized charge (fermion mass).
In perturbation theory $\sigma_{\rm Re}({\mathcal O}) = O(e_R^2)$ and 
$\langle \cos\pi\eta_a \rangle_{\rm Re} = 1 + O(e_R^2)$. 
From this and (\ref{estimate1}) it then follows that in the chiral limit
\begin{equation}
\langle {\mathcal O}\rangle \rightarrow \langle {\mathcal O}\rangle_{\rm Re}
\rightarrow  \langle {\mathcal O}\rangle_0,
\end{equation}
where the subscript `0' denotes the result of the free theory.
This would mean that the chiral theory is trivial, like massless QED. A 
direct estimate 
of $\langle {\mathcal O} \cos\pi\eta_a \rangle_{\rm Re}^{\rm con}$ leads
to the same conclusion. As far as we can see, our argument could go 
wrong only if $\eta_a$ is not described by perturbation theory and 
$\langle \cos\pi\eta_a \rangle_{\rm Re} \approx 0$.

\section{Conclusions} 

We have presented first results of the effective action of the chiral U(1)
theory on small lattices, which led us to argue that the theory is trivial. 
To substantiate our results we obviously need to do calculations on larger 
lattices and at further couplings. If true, our results would perhaps explain
why neutrinos are massiv, in the same way as QED tells us that electrons
must be massiv.

\vspace*{0.35cm}

We thank A. Thimm for collaboration in the early stages of this work.

\end{document}